\documentclass{pj}
\usepackage{graphicx} 
\usepackage[export]{adjustbox}
\usepackage{doi}
\usepackage{siunitx}
\usepackage{amsmath}
\usepackage{booktabs}
\usepackage{tabularx}
\usepackage{microtype}

\usepackage{color}
\usepackage{multicol}

\definecolor{tbl-background}{RGB}{255,251,243}
\newcommand{\tablesectionheader}[1]{\textbf{#1}}

\fancypagestyle{firstpage}{
    \fancyhead{}
    \fancyfoot{}
    \fancyfoot[C]{\small\thepage}
}

\fancypagestyle{main}{
  \fancyhead{}
  \fancyfoot{}
  \fancyfoot[C]{\small\thepage}
}
\definecolor{sectioncolor}{RGB}{133,116,55}
\definecolor{figtablabel}{RGB}{133,116,55}
\definecolor{titlecolor}{RGB}{0,48,87}

\begin{document}

\title{Equivalent Circuit Modeling and Design of a Reconfigurable Loaded Dogbone Metasurface Element}
 
\author[1,2]{Christopher~T.~Howard}
\author[2]{William~D.~Hunt}
\author[1,2]{Kenneth~W.~Allen}

\shortauthor{Howard \textit{et\,al.}}
\affil[1]{Advanced Concepts Laboratory, Georgia Tech Research Institute, Atlanta, GA 30318 USA}
\affil[2]{School of Electrical and Computer Engineering, Georgia Institute of Technology, Atlanta, GA 30332}

\emails{Corresponding authors:~Christopher~Howard (christopher.howard@gtri.gatech.edu).}

\StartPage{1}
\volume{xxxx}{xx}{xxx}{xx}
\paperdoi{xxxxxxxxxxxxxx} 
\history{Date Month Year}{}{xxx}{xxx}

\begin{Abstract}
The accelerating trend of active metasurfaces -- such as those incorporating non-Foster matching, programmable control, or space-time modulation -- adds complexity to the computational electromagnetic (CEM) simulation landscape. In this work, we present an equivalent circuit model (ECM) for a particular periodic array element -- the dogbone element -- that incorporates an arbitrary sub-cell electronic circuit. The ECM is capable of predicting the overall surface impedance with any changes in the circuit's impedance. Three full-wave CEM simulations fully determine the EC, and the scattering parameters computed from the ECM for an embedded circuit, however complex, should be substantially similar to those produced by the full-wave code. With metasurfaces for which it is possible to obtain an ECM by this  method, the reflection or transmission behavior of the surface under varying circuit conditions can be computed quickly without need of any further full-wave solutions. After highlighting the importance of fully characterizing parasitics in embedded components, we apply the method to the design of a varactor-based tunable metasurface, which upon fabrication and focused beam measurement demonstrates excellent model-measure agreement.
\end{Abstract}

\section{Introduction}

The inclusion of active electronic circuitry directly into electromagnetic structures is one defining aspect of the newest generation of electromagnetic engineering, as it enables antennas and metasurfaces to exceed many fundamental limitations, such as reciprocity \cite{wang2023} and bandwidth constraints \cite{kalmykov2021}, which assume that the structures are entirely passive. In addition, the growing field of time-varying structures, including tunable frequency-selective surfaces and reconfigurable intelligent surfaces (RISs), allows static electromagnetic structures to produce unusual responses, such as apparent motion \cite{ramaccia2020}, information encoding \cite{zhang2018}, spatial analog computing \cite{zangeneh-nejad2021}, and more, which have applications ranging from autonomous vehicular communication \cite{zhu2022b} to next generation wireless networks \cite{pei2021a,mizmizi2024, ashraf2023}.

Modeling these structures is challenging. The electronics involved are generally much smaller than the structure being modeled in a full wave simulator; their impact on the electromagnetic response of the structure cannot be described by varying material properties in space, as is usually the underlying description of the system used in CEM codes. In addition, the underlying physical models describing current-voltage behavior of the semiconductors and other circuit components requires modeling across multiple domains. There are, in fact, analytical solutions for the impact of some linear impedance -- resistance, capacitance, or inductance -- in a particular region of space, which leads to the so-called lumped element boundary conditions available in many commercial solvers. However, we note that while these boundary conditions are useful for many applications such as unit cell miniaturization \cite{khuyen2017}, they are inflexible for the modeling of active, sometimes nonlinear circuits which are used in reconfigurable and time-varying applications.

\begin{figure*}[!t]
    \includegraphics[width=6in,center]{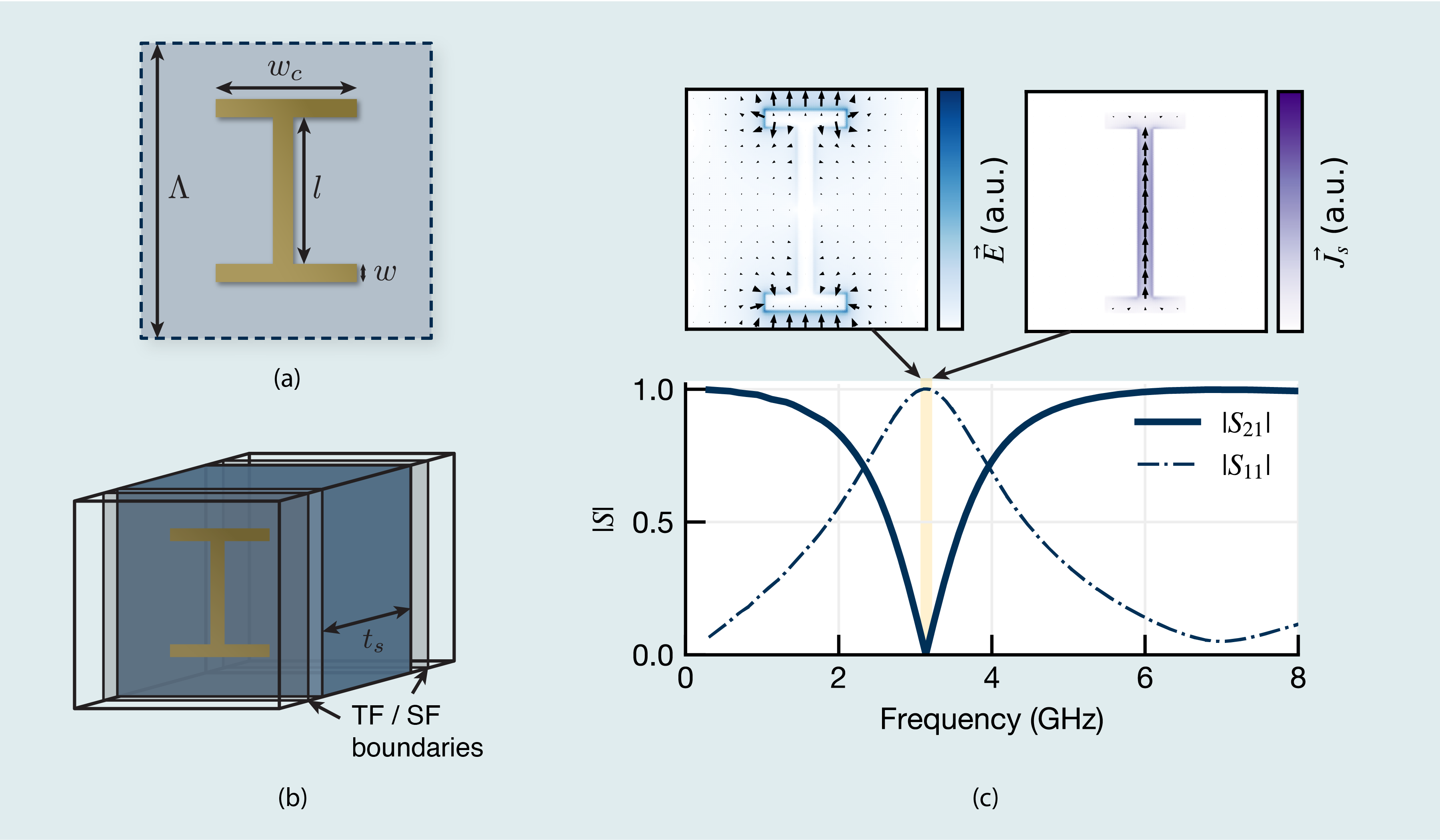}
    \caption{Depiction of the end-loaded dipole ("dogbone") surface, showing (a) dimensional parameters of the surface, (b) a nominal geometry for a periodic FDTD model, and (c) a nominal single-resonance frequency response and electric field and currents on the surface for a particular configuration of the surface ($l = \SI{21.34}{\mm}$, $w = \SI{2.03}{\mm}$, $w_c = \SI{10.16}{\mm}$, $\Lambda=\SI{30.48}{\mm}$, $\epsilon_{rs}=3.0$, $t_s = \SI{1.524}{\mm}$).}
    \label{fig:dogbone-fss}
\end{figure*}

One general approach may be found in the use of hybrid simulators; that is, codes which simultaneously use a full wave electromagnetic solver to model the overall structure's response to incident electromagnetic energy and a circuit simulator to model the sub-wavelength circuit components of the structure. This generally involves sampling the fields at a particular location in space and injecting them into a transient step of a circuit solver such as SPICE \cite{pederson1984}, and is therefore most suitable for time domain methods such as those using finite differences (FDTD) \cite{thomas1994}, finite elements (FETD) \cite{he2010}, or integral equations (TDIE) \cite{bagci2005}. Unfortunately, full-wave modeling approaches are unwieldy for modeling electromagnetic structures which contain active electronic circuits which might perturb the electromagnetic response in a variety of ways; a large number of simulations may necessary to fully characterize the device across the entire domain of its operating conditions. A sufficiently detailed computational model of the structure may take many minutes or hours to solve for a single static case, and the modification of any one parameter requires the model to be solved again. For structures with only one tunable parameter, a simple parametric sweep may be used to characterize the device, but for models with multiple tunable parameters, this quickly becomes intractable.

In this work, we use only three full-wave simulations to obtain an equivalent circuit network for a dogbone metasurface which can then be used to predict metasurface performance over a wide range of lumped impedance conditions. While lumped element boundary conditions in commercial solvers enable the embedding of discrete RLC components, the equivalent circuit approach allows the embedding of circuits of arbitrary complexity. For example, the dogbone metasurface considered in this work has a variable resonant frequency enabled by varactor diodes; a relatively complex Foster's equivalent circuit describing the small-signal performance of the varactor including parasitics is obtained and incorporated into the metasurface equivalent circuit. The results from the equivalent circuit are compared to those obtained from a hybrid FDTD-SPICE co-simulator. We explore a particular design prepared with this technique and show good agreement between the equivalent circuit model and measured transmission through the dogbone metasurface.

\section{Modeling of dogbone periodic array with lumped element}

To explore the impact of lumped components on metasurface impedance, we select a metasurface which has a simple enough resonant response to be represented by an equivalent circuit but with enough parametric diversity create a wide range of surface impedance conditions. The dogbone shape, or end-loaded dipole, shown in \autoref{fig:dogbone-fss}(a), has a single resonance with a band-stop response similar to that of a simple dipole array, but with a more compact shape for a given resonant frequency \cite{xu2018}. The added degree of freedom provided by the end loading allows the designer to vary the amount of capacitive versus inductive reactance contributing to the overall surface impedance. The parameterized dimensions of the surface are depicted in \autoref{fig:dogbone-fss}(a). The dipole has length $l$, end-loading width $w_c$, and trace width $w$, centered on a unit cell with square periodicity of $\Lambda$. 

The full-wave results generated throughout this article are obtained from a finite-difference time-domain (FDTD) code internally developed at GTRI, with a typical model geometry depicted in \autoref{fig:dogbone-fss}(b). The model is discretized onto an FDTD grid with cell size $\Delta$, and for every FDTD result in this work, $\Delta=\SI{0.254}{\mm}$, which is approximately $\lambda/120$ for the highest frequencies under consideration.

A hybrid FDTD-SPICE co-simulator is used to simulate the incorporation of lumped elements. The method for incorporating SPICE into the Yee algorithm is described in \cite{taflove2005}; current is sampled at each time step on a particular Yee cell edge -- the location of the SPICE port -- according to Ampere's Law,
\begin{equation}
    I_N(t) = \oint_C \vec{H} \cdot d\vec{C}
\end{equation}
where $C$ is a closed contour around the location of the the SPICE port, obtained by sampling the magnetic field at each of the neighboring Yee cells. This value is used as the current source in a Norton equivalent circuit in SPICE looking into the the FDTD space lattice, which has a parallel intrinsic capacitance of $C_{\text{lattice}} = 3\varepsilon_0 \Delta$ where $\varepsilon_0$ is the permittivity of free space and $\Delta$ is the Yee cell size \cite{taflove2005}. A transient SPICE simulation is run for a single time step, and the resulting port voltage is used to update the electric field in the FDTD grid.

\section{Equivalent circuit modeling of dogbone array}

\begin{figure}[t!]
    \includegraphics[width=2in,center]{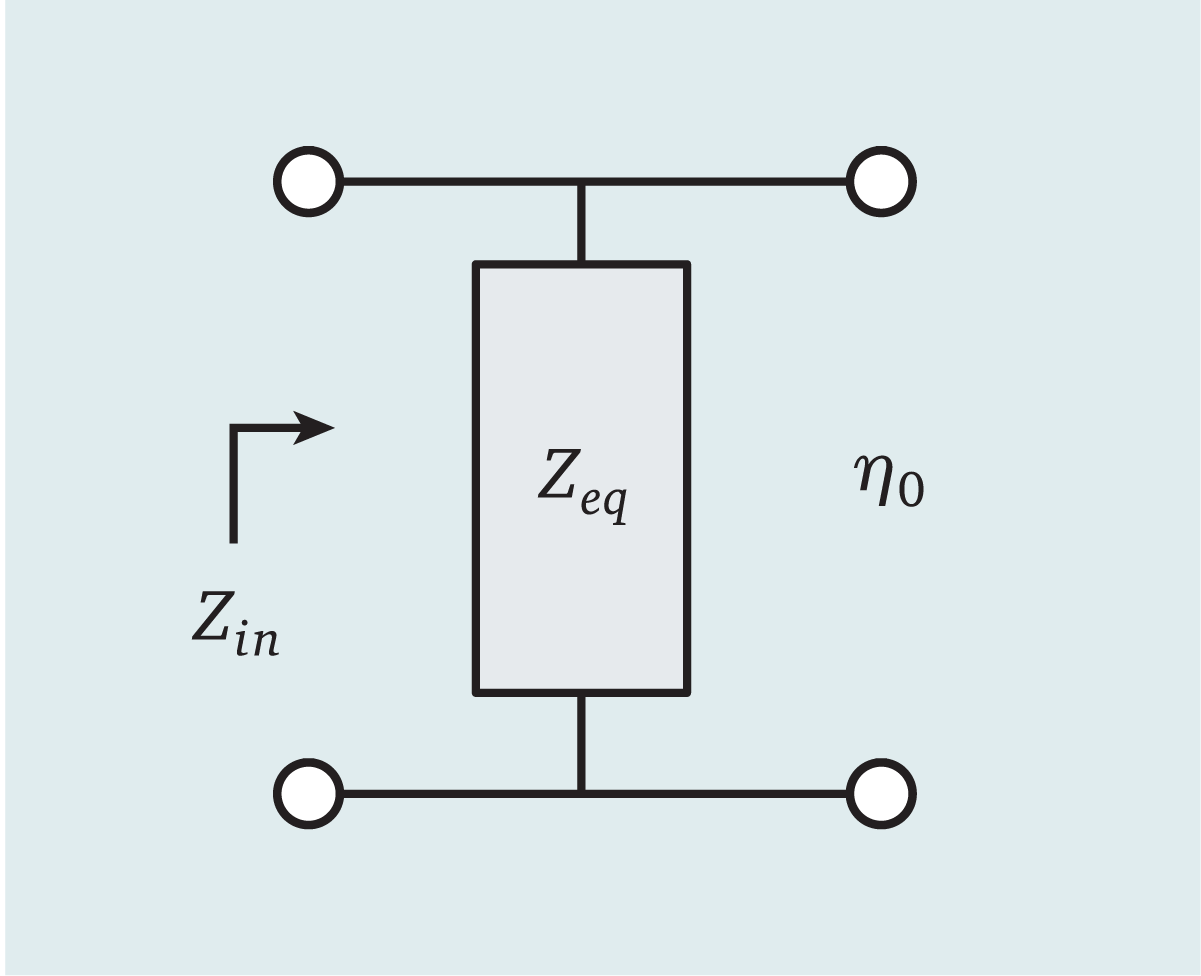}
    \caption{Equivalent circuit modeling represents metasurface behavior as an two-port impedance network $Z_{eq}$ terminated by the free-space impedance $\eta_0$.}
    \label{fig:simple-ecm}
\end{figure}

Equivalent circuit modeling (ECM) is a useful way of capturing the complex frequency-dependent behavior of metasurfaces in only a few parameters. In principal, it suggests that a metasurface behaves as a sheet impedance, which can be represented by an admittance $1/Z_{eq}$ as depicted in \autoref{fig:simple-ecm}; this admittance may be further decomposed into some combination of passive components. ECM is especially useful for metasurface design because it provides an intuitive basis for mapping physical features of a periodic surface to its frequency spectra; continuous current paths provide inductance to the circuit and discontinuities capacitively accumulate charge. The ECM method is not merely a curve-fitting exercise; the equivalent circuit method represents a rigorous modal decomposition of the fields at the surface, where each mode is approximated as an inductor (TE modes) or a capacitor (TM modes) at low frequencies \cite{guglielmi1989}, otherwise known as the homogenization condition \cite{costa2014}. The ECM technique thus represents a robust solution to Maxwell's equations for planar periodic structures, but only when the wavelengths of interest are substantially larger than the periodicity of the structure. At higher frequencies, the characteristic admittances of the lower-order modes become highly dispersive and can no longer be suitably represented by a single lumped component, although this range can be extended somewhat by decomposing each admittance into a number of components according to its Taylor series expansion \cite{dubrovka2006} or through carefully computed correction factors \cite{lee1985}.

\subsection{Lumped components in equivalent circuit}

\begin{figure}[t!]
    \includegraphics[width=3.5in,center]{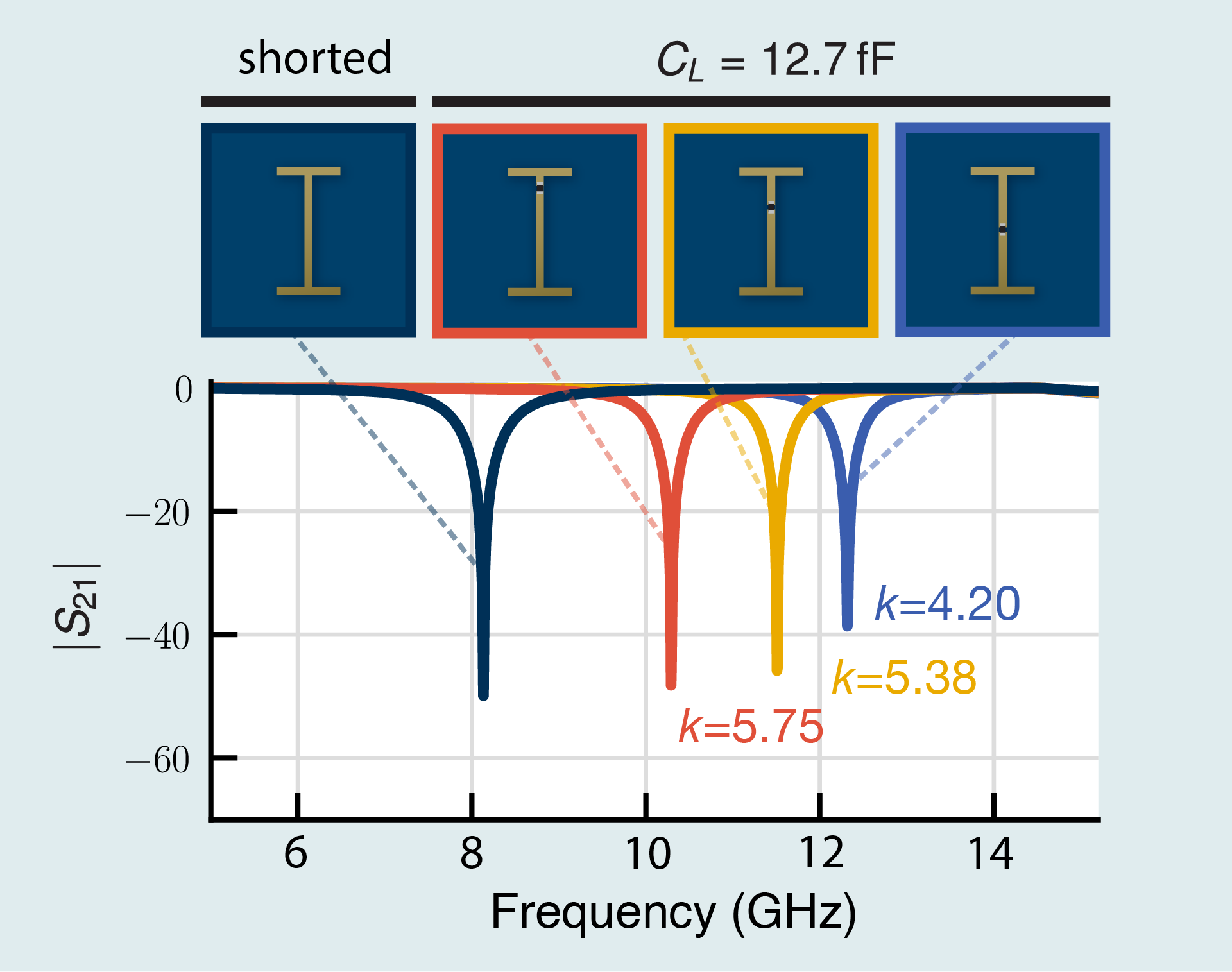}
    \caption{The impact of of a lumped capacitor ($C_L = \SI{12.7}{\pF}$)in a metasurface element varies with capacitor location, with the effect captured by the capacitance scaled by a constant $k$.}
    \label{fig:offset-lumped-comparison}
\end{figure}

The incorporation of circuit components into the ECM method is complicated by a lack of obvious correlation between a component's placement on the periodic surface and its location within the circuit topology. For example, the intuitive representation of a simple dipole array is that of a series LC resonator, and a lumped capacitor may be placed across a gap introduced anywhere along the length of the conductor. One might argue that any current flow along the inductive element passes through the component, and as such, the lumped capacitor is placed in series with the LC elements in the equivalent circuit representation. Such a strategy has been shown to work in limited cases, such as for a centered connection between patch elements as in \cite{costa2021a} or fully connected dipoles \cite{zhao2019}. Lumped components have been directly inserted into equivalent circuit models for other metasurface designs \cite{zhao2020,bayatpur2009}, but without measured or full-wave data, it is difficult to assess the validity of the assumption. In \cite{guo2019}, lumped capacitors placed directly into the ECM required changes in the surface equivalent circuit values in order to obtain agreement with full-wave modeling. We should not expect that we can move a lumped component to any arbitrary location on the metasurface and expect that its contribution to the equivalent circuit remain unchanged; for example, this would suggest that the effect induced by the component is invariant to its location along a dipole metasurface element, which is not the case, as illustrated in \autoref{fig:offset-lumped-comparison}.

\begin{figure*}[ht!]
    \includegraphics[width=6in,center]{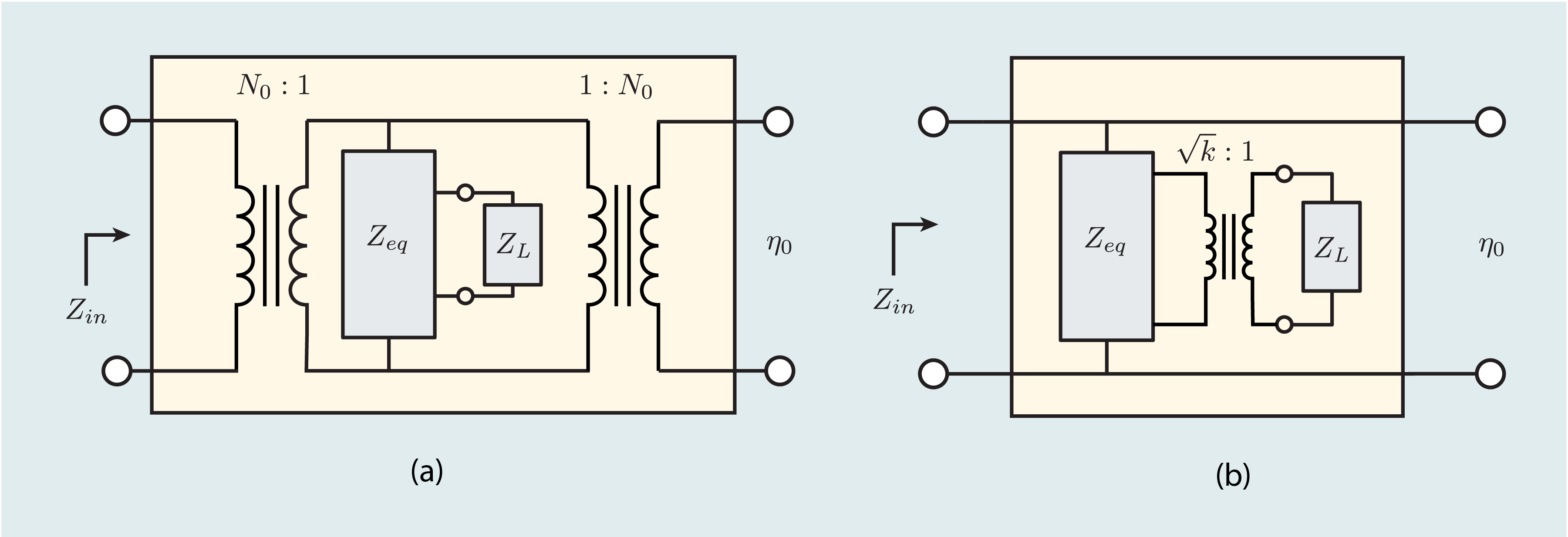}
    \caption{Two different (but topologically identical) representations of an equivalent circuit model incorporating lumped circuit components.}
    \label{fig:overall-ecm-transformer}
\end{figure*}

The lumped component representation in in the equivalent circuit must depend on its spatial location. 
We should be reminded that the ECM is a \textit{homogenized representation} of complex electromagnetic fields produced by currents which vary over the surface; the effective inductance of the surface, for example, is a single value integrating the effects of infinitesimal elements of current across the surface conductors. The amount of current which actually flows through the component relative to the total surface current on the conductors is related to the magnitude of current that would exist at that location in the absence of a component altogether. The multi-modal equivalent circuit of \cite{hum2017} accommodates an arbitrary lumped component of impedance $Z_L$ by noting that the component affects how effectively an incident wave is coupled into currents on the surface; that is, the location of the lumped component perturbs the spatial current profile on the surface, which in the multi-modal formulation, determines the resulting surface impedance. This is captured in the equivalent circuit model by input and output transformers with turn ratios of $N_0$ to $1$, as shown in \autoref{fig:overall-ecm-transformer}(a). Unfortunately, this robust formulation fails to isolate the component from a static equivalent circuit; varying its spatial location changes both the surface impedance $Z_{eq}$ and the turn ratio $N_0$, with the strongest coupling and therefore highest $N_0$ value occurring when the component is placed at a peak in the current spatial profile.

A more intuitive representation, from \cite{luther2013}, notes that the impedance contribution from an embedded component is related to the potential energy of the fields at that location. The effective impedance contribution $Z_{L,\text{eff}}$ from a component with impedance $Z_L$ inserted into the circuit is simply scaled by a constant $k$, yielding
\begin{equation}
    Z_{L,\text{eff}} = k Z_L \quad \text{for} \quad k \geq 1
\end{equation}
where $k$ is determined by the location of the lumped capacitor. The generalized circuit for this concept is shown in \autoref{fig:overall-ecm-transformer}(b). The two circuit topologies in \autoref{fig:overall-ecm-transformer} are identical in that they produce the same impedance spectra, but have different scaling applied to the impedance values of the constituent components. We will refer to the latter throughout this article, as it more cleanly separates the intrinsic impedance of the metasurface from the lumped components added to it.

The equivalent surface impedance $Z_{eq}$ is therefore a two-port network loaded with the scaled lumped component $k Z_L$ on its output port. The next section turns to the development of a method for determining the network topology for a particular metasurface shape.

\subsection{Method for extraction of equivalent circuit impedance}

In theory, for an equivalent circuit containing some number of components, an equal number of individual scattering data points are necessary to fully characterize the circuit. For example, it is possible to obtain the component values for an equivalent circuit consisting of an inductor and a capacitor from a frequency domain solver with simulations of reflection at two different frequencies. The reality is however that without \textit{a priori} knowledge of the resonant frequency of the surface, it is difficult to choose appropriate sampling frequencies to generate a well-conditioned system of equations for this technique. The FDTD algorithm generates a large number of frequency points from a single simulation and thus forms an over-determined system of equations; solving for the component parameters using least-squares or some iterative method offers some defense against this problem and is able to inspire confidence in the quality of fit over a large frequency range.

The first step in the process is obtaining the equivalent circuit impedance from the reflection data produced by the full wave solver
via
\begin{equation}
    Z_{eq} = -\eta_0 \frac{1 + S_{11}}{2 S_{11}}.
    \label{eq:Zeq-from-S11}
\end{equation}
The reflection coefficient $S_{11}$ must be reflection from the surface alone, deembedded from the effects of any substrates, which appear as transmission lines on either side of the surface.
\newcommand{\Zsurf}{Z_{\textit{surf}}}
\newcommand{\jw}{j \omega}
The equivalent impedance obtained from the reflection data represents the impedance seen looking into an impedance network of unknown complexity. We are left with the task of determining how an embedded circuit within that network affects the electromagnetic response. As discussed previously, it is not sufficient to incorporate an embedded circuit by simply adding the impedance of the components to the surface impedance; here we also note that it does not suffice to just add a scaled version of that impedance. Rather, the added component represents a load on an arbitrary equivalent network, the inner topology of of which must be determined through either some general approach (similar to unterminating a measurement fixture \cite{bauer1974}) or through informed guesswork (otherwise known as intuition).

\begin{figure*}[ht!]
    \includegraphics[width=6in,center]{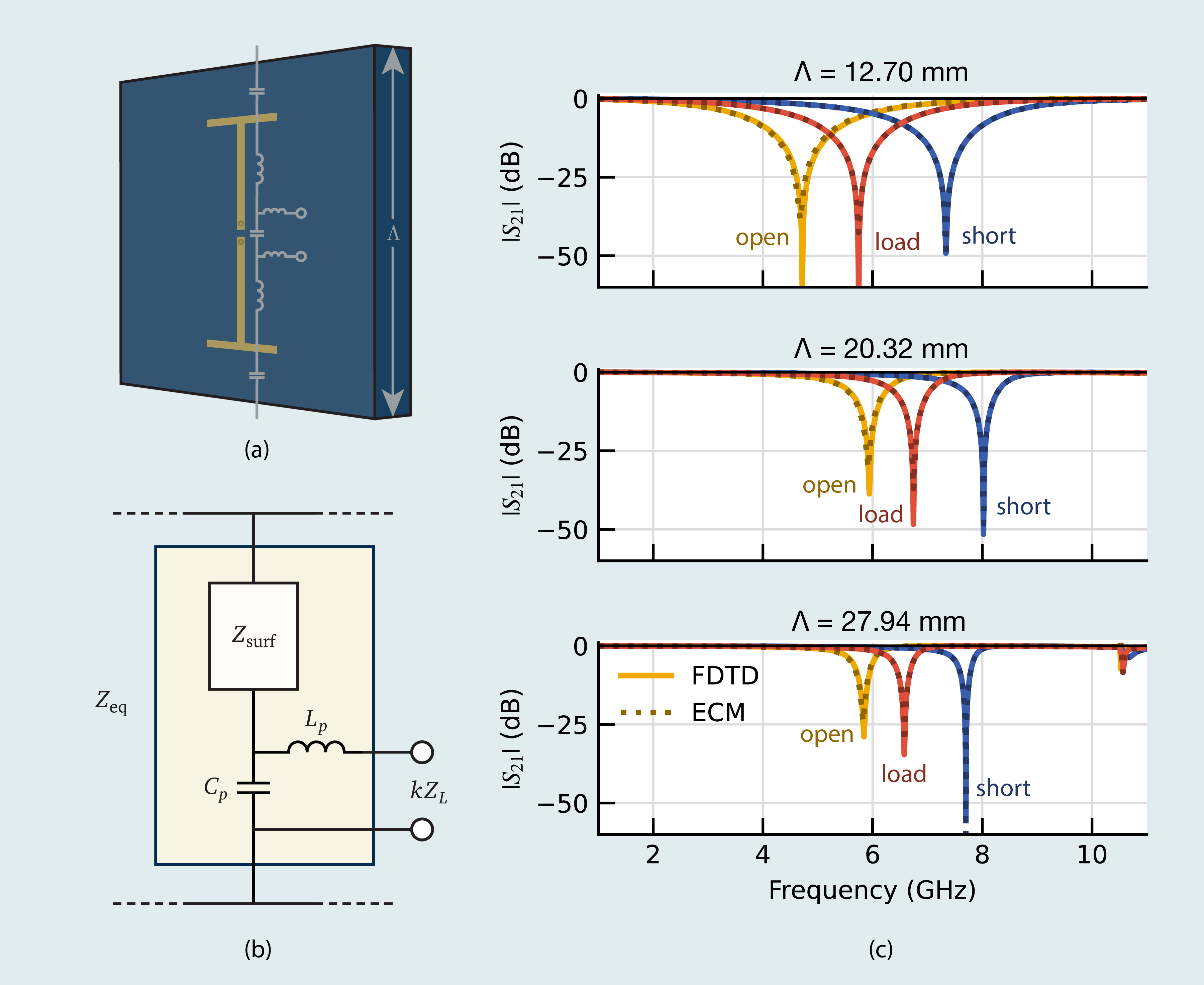}
    \caption{(a) Inspection of metasurface geometry provides a reasonable equivalent circuit. (b) The equivalent circuit representation of the metasurface with lumped component impedance $Z_L$. Transmission obtained for open, short, and load cases for a dogbone array with parameters $l = \SI{11.40}{\mm}$, $w = \SI{0.508}{\mm}$, $w_c = \SI{4.572}{\mm}$, $g=\SI{0.254}{\mm}$, $\epsilon_{rs}=3.0$, $t_s = \SI{1.524}{\mm}$ and three different periodicities $\Lambda$.}
    \label{fig:ecm-fit}
\end{figure*}

The dogbone surface depicted in \autoref{fig:dogbone-fss} is modified to accommodate a lumped component as shown in \autoref{fig:ecm-fit}(a). The element is split in the center and vias used to route current to pads on the backside of the substrate, where the lumped component sits. 
The impedance response of this surface is primarily determined by the geometry of the surface's conductive elements, which represents an impedance contribution $\Zsurf$ that remains unperturbed by the addition of any lumped components. While this impedance could certainly be decomposed into individual components for the circuit analysis, it is not necessary to do so, and there is some advantage in not doing so. At a large periodicity $\Lambda$, the dispersive nature of the evanescent modes is poorly approximated by an ideal capacitor or inductor, and by extension, the impedance curve is poorly approximated by a lumped equivalent circuit model. Instead, we seek to represent as much of the equivalent circuit as possible as an arbitrary impedance element $\Zsurf(f)$ which varies with frequency, and a minimal number of components are added to the circuit insofar as they clarify the relationship between $\Zsurf$ and the lumped component $Z_L$. It is also possible to to use full-wave simulations to determine an arbitrary two-port impedance network relating surface impedance to the circuit port \cite{chou2021, zhang2022, zhang2024a, oconnor2019, ramachandran2022}, but doing so reveals little intuitive insight into the nature of the coupling between surface and component; the computed network remains a "black box". The method used to obtain an equivalent circuit in this work is not dissimilar from these approaches, but seeks to qualify the relationship between standalone metasurface and any included lumped components or active circuitry. This middle road between arbitrary network and complete equivalent circuit formulation yields intuitive explanation for the limitations of tunability of the reconfigurable metasurface.

Inspection of the geometry allows for some anticipation of a reasonable equivalent circuit model, as shown in \autoref{fig:ecm-fit}(b). The impedance is dominated by the series combination of the inductive dipole element and capacitive coupling between the ends of the dipole, which is captured as $\Zsurf$. To accommodate a lumped component in the dipole, the continuous current path is severed, creating a capacitive gap represented as $C_p$ in parallel with the lumped element. The vias routing current to the backside of the substrate contribute a small inductance, denoted as $L_p$, in series with this load. This equivalent circuit has the total surface impedance
\begin{equation}
    Z_{eq} = \Zsurf + \frac{1}{\jw C_p} \parallel (\jw L_p + k Z_L)
\end{equation}
for some component with impedance $Z_L$. 

While the topology of the circuit may be obtained through physical intuition, the magnitudes of each component are more difficult to precisely define. Simple curve fitting of the equivalent circuit model to a single full-wave simulation is prone to uniqueness issues; the values obtained for a particular loading condition may produce a very good fit but fail to properly predict the perturbation of the circuit by other loads. We instead seek to rigorously determine the contents of the equivalent circuit through a network de-embedding. Three separate full-wave simulations are performed, with the location of the lumped component in the circuit replaced in turn by an open, a short, and a known lumped component. The open ($Z_L \to \infty$) and the short ($Z_L \to 0$) cases are useful to illustrate the theoretical range of tunability under varying capacitor; any possible capacitor must have a reactance between these two values. The impedance for all three simulations can be respectively extracted with \autoref{eq:Zeq-from-S11} and then formulated into a system of equations:
\begin{equation}
\begin{cases}
    Z_{\textit{open}} = \Zsurf + \frac{1}{\jw C_p} \\
    Z_{\textit{short}} = \Zsurf + \frac{1}{\jw C_p} \parallel \jw L_p  \\
    Z_{\textit{loaded}} = \Zsurf + \frac{1}{\jw C_p} \parallel (\jw L_p + k Z_L).\label{eq:Zloaded}
\end{cases}
\end{equation}

First, the values for $C_p$ and $L_p$ are obtained by subtracting $Z_{\textit{short}}$ from $Z_{\textit{open}}$ and fitting the resulting data over the entire frequency range using nonlinear least-squares according to
\begin{align}
    Z_{\textit{open}} - Z_{\textit{short}}
        &= \frac{1}{\jw C_p} - \frac{1}{\jw C_p} \parallel \jw L_p \\
        &= \frac{1}{\jw C_p (1 - \omega^2 L_p C_p)}.
\end{align}

Once, $C_p$, $L_p$ and $\Zsurf$ are obtained from this curve fit, a lumped component with impedance $Z_L$ is introduced into the model in order to obtain the coupling term $k$. Any resistive or reactive component will do, but its value must be thoughtfully chosen to have an impedance magnitude in the neighborhood of the impedance of $C_p$ to ensure that the load sufficiently perturbs the overall impedance. A capacitive load of $C_L = \SI{0.30}{\pF}$ is modeled for the loaded case using lumped element boundary conditions. Non-linear least squares is used to fit \autoref{eq:Zloaded} to find $k$. The values of for $C_p$, $L_P$, and $k$ for the same dogbone surface with varying periodicity $\Lambda$ of \SI{12.70}{\mm}, \SI{20.32}{\mm}, and \SI{27.94}{\mm} are listed in \autoref{tbl:extracted-circuit-parameters}, and a comparison of the fit of transmission versus frequency derived from the ECM compared to that from FDTD for each case is shown in \autoref{fig:ecm-fit}.

\begin{table}
    \centering
    \caption{Extracted ECM parameters for dogbone surface in \autoref{fig:ecm-fit}}
    \label{tbl:extracted-circuit-parameters}
    \colorbox{tbl-background}{\begin{tabular}{lccc}
                                    & $L_p$ (nH)    & $C_p$ (fF)        & k \\\midrule
         $\Lambda = \SI{12.70}{\mm}$&  0.637        &  109              & 1.43 \\
         $\Lambda = \SI{20.32}{\mm}$&  1.92         &  39.1             & 3.86 \\
         $\Lambda = \SI{27.94}{\mm}$&  4.03         &  20.8             & 7.13 \\
    \end{tabular}}
\end{table}

\subsection{Prediction of equivalent impedance from component variation}

\begin{figure*}[ht!]
    \includegraphics[width=6in,center]{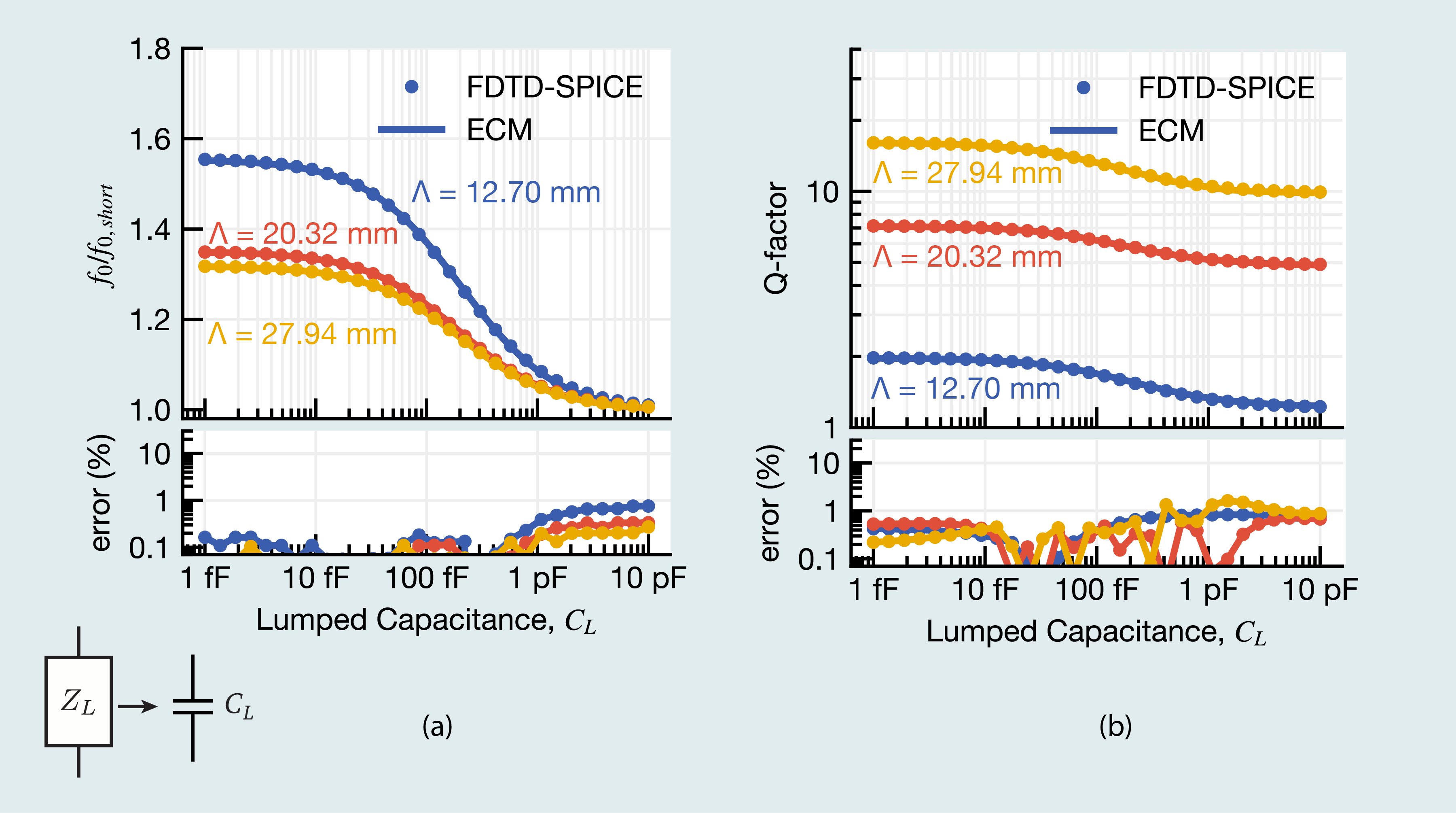}
    \caption{Comparison of FDTD-SPICE versus ECM in computing the resonant frequency for the models from \autoref{fig:ecm-fit}. }
    \label{fig:f-vs-Cd}
\end{figure*}

The resulting ECM can then used to predict scattering of the surface for any possible load. A capacitance which varies between \SI{1}{\fF} and \SI{10}{\pF} is supplied as a load to the ECM, and from the resulting scattering parameters the resonant frequency and Q-factor of the fundamental resonance is obtained. The curves produced by the model are compared to FDTD-SPICE simulations in \autoref{fig:f-vs-Cd}. They approach limits on either side of the capacitance range as a result of the load capacitance being placed in parallel with the parasitic capacitance $C_p$. Low capacitance values provide a large negative reactance at the frequencies of interest, which appears as an open, and current primarily flows through the parasitic capacitance $C_p$. Similarly, high capacitance values cause the reactance to approach zero, which is functionally identical to the shorted case. We would therefore expect that the resonant behavior is bounded between the shorted case and the open case when the reactance of the load is negative. 

\begin{figure*}[ht!]
    \includegraphics[width=6in,center]{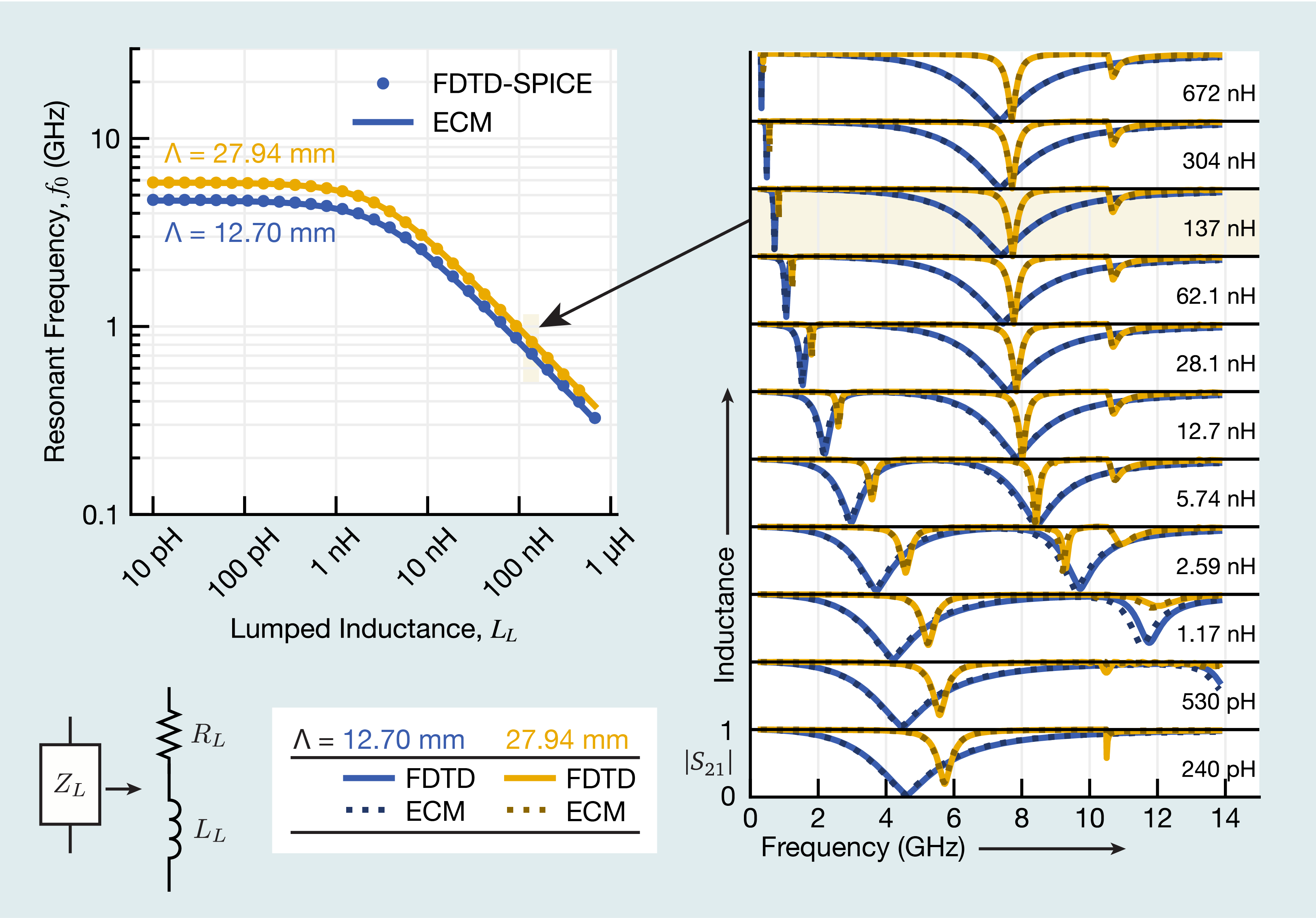}
    \caption{Comparison of resonant frequency obtained from FDTD-SPICE simulation and ECM prediction for a range of inductance values in series with a \SI{4}{\ohm} resistor. The figure on the right shows agreement in transmission between FDTD-SPICE results (solid lines) and the ECM results (dashed lines) for a sampling of the inductance values plotted on the left.}
    \label{fig:inductance-ecm-model}
\end{figure*}

To further illustrate the general predictive power of the equivalent circuit model, the capacitive load is replaced by an arbitrary inductive load $L_L$ with series resistance $R_L=\SI{4}{\ohm}$, and scattering parameters predicted from the ECM. The inductance is varied between \SI{10}{\pH} and \SI{1}{\uH}. Comparison between the ECM prediction and full-wave FDTD-SPICE simulation is shown in \autoref{fig:inductance-ecm-model}. When the added inductance is lower than the via inductance $L_p$ (on the order of \SI{1}{\nH}), little perturbation occurs, and the scattering response is practically identical to the shorted case. However, a larger inductance depresses the fundamental resonances towards zero frequency without limit (this analysis ignores the relative improbability of generating such a large inductance within the volume prescribed). In addition, a second resonance, normally well within the diffraction regime, descends in frequency to a limit of about \SI{8}{\GHz}, and is properly predicted by the ECM model.

\section{Small-signal modeling of varactor diode}

\begin{figure}[htbp!]
        \includegraphics[width=3.5in,center]{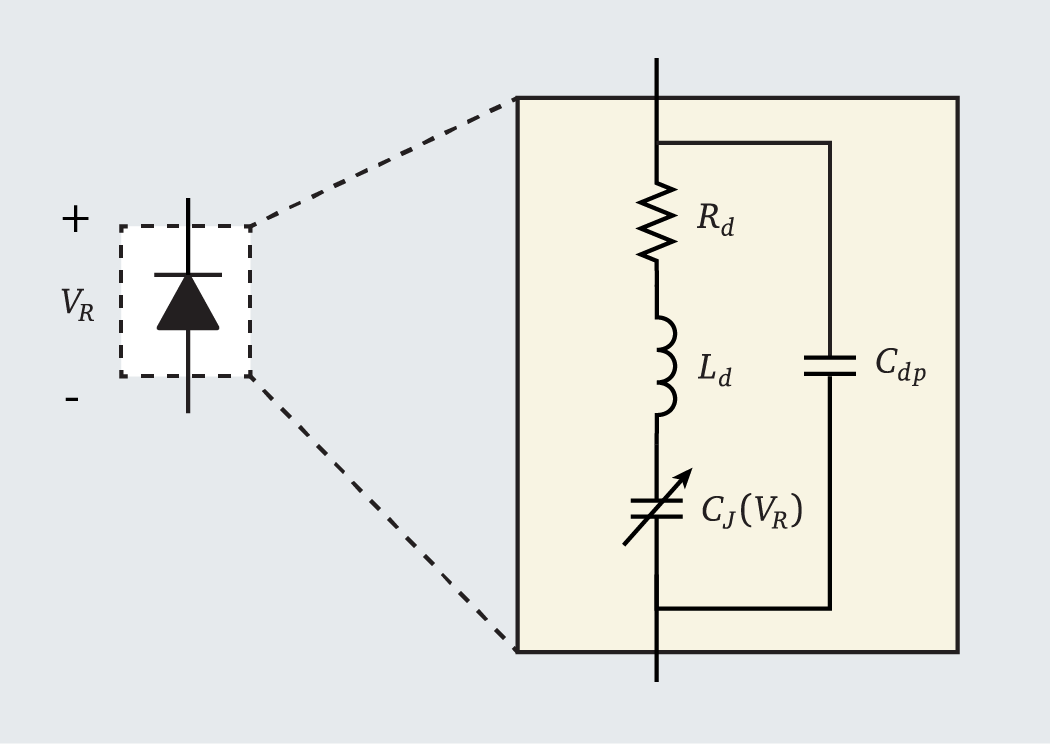}
        \caption{A typical small-signal representation of a diode consists of a variable junction capacitance $C_J$ dependent on applied reverse bias voltage $V_R$, and a number of other parasitic components contributing to the overall impedance \cite{rudyk2021}.}
        \label{fig:diode-small-signal}
\end{figure}

With confidence that the ECM is capable of properly predicting scattering from a surface loaded with any arbitrary impedance, it is now possible to turn to more relevant loads for establishing tunable control over the resonant frequency of the surface. Varactors are diodes designed to have strong variation in junction capacitance relative to applied reverse bias voltage. While in general this means that a diode has a complex nonlinear voltage-current relationship, it is sufficient for low-power applications such as this one to use a small-signal approximation of the diode consisting only of lumped passive components \cite{rudyk2021}. One such representation is shown in \autoref{fig:diode-small-signal}. The overall capacitance of the diode is primarily determined by the junction capacitance $C_J$, which varies with reverse bias voltage $V_R$, but this idealized capacitance is affected by the presence of parasitics in the diode. Some inductance $L_d$ is caused by the traces and packaging of the diode, which also contribute some resistance which can be combined with any junction resistance and represented by $R_d$. Finally, there is a parallel package capacitance $C_{dp}$.

Accurate modeling of a metasurface with an embedded small-signal diode representation requires a high-quality source of data for both the junction capacitance $C_J$ and the values of the parasitic elements. These values are sometimes available from the manufacturer. However, they are often characterized at low frequencies (e.g. \SI{1}{\MHz}) at which many of the important parasitics at microwave frequencies cannot be resolved.

For this work, a GaAs hyperabrupt varactor diode made by Macom (MAVR-011020-1411 \cite{mavr011020}) was selected, primarily for its low junction capacitance. For varying reverse bias of 0V to 15V, this diode presents a series junction capacitance $C_J$ of \SIrange{0.036}{0.23}{\pF} with a series resistance of \SI{13.2}{\ohm}. The datasheet values for this capacitor have a large range of variability; for example, at the zero-bias condition, the junction capacitance $C_{J0}$ has up to 44\% variation (\SIrange{0.19}{0.275}{\pF}). Of note is that the datasheet lists this as the total capacitance of the varactor, implying that it includes any package capacitances, but it is consistent with the SPICE models provided by Macom and the relatively low test frequency of \SI{1}{\MHz} that this is in fact the junction capacitance. Relevant datasheet values and their ranges are reproduced in \autoref{tbl:macom-datasheet-values}. The datasheet also specifies an equivalent circuit for the diode that is identical to that of \autoref{fig:diode-small-signal}, but provides no component values for that model.

\begin{figure}[ht!]
    \includegraphics[width=3.5in,center]{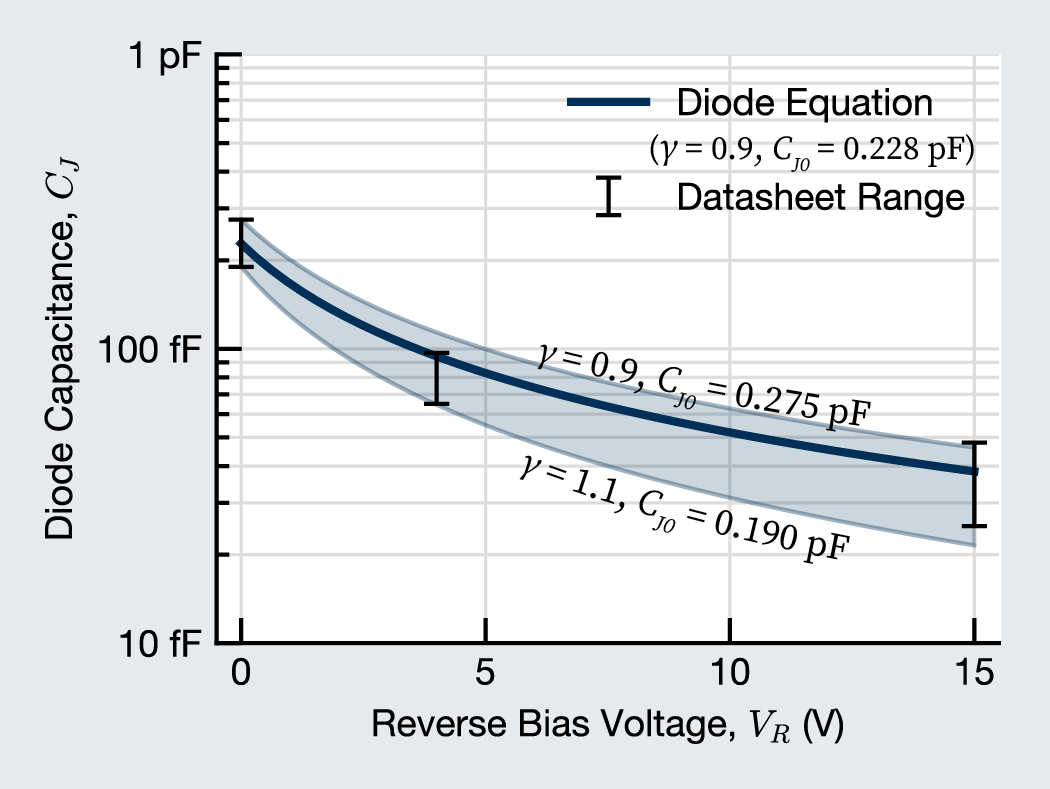}
    \caption{Junction capacitance of the MAVR-011020-1411 varactor diode based on reverse bias voltage. The dark line represents a typical C-V relationship, while the shaded region represents the range of allowable values based on the datasheet tolerances. Also provided by the datasheet are capacitance ranges for 0, 4, and 15 volts, annotated by the dark error bars.}
    \label{fig:diode-cv}
\end{figure}

The junction capacitance value is only available from the datasheet for a few discrete voltages, but we need to understand the capacitance behavior over the entire bias voltage range. This capacitance-voltage relationship is defined by the reverse-bias capacitance equation
\begin{equation}
    C_{J}(V_R) = C_{J0} \left(1 - \frac{V_R}{V_J}\right)^{-\gamma}
    \label{eq:diode-cv-equation}
\end{equation}
where $V_J$ is the built-in junction potential of the diode (which the MACOM model somewhat inexplicably specifies to be \SI{2.4}{\volt}) and $\gamma$ is the slope of the $\log C$ vs. $\log V$ curve \cite{varactor2008}. The expected relationship between junction capacitance and reverse bias voltage is shown in \autoref{fig:diode-cv}. Because of the wide range of datasheet values provided, a region of possible capacitance values is plotted based on the minimum and maximum values of $C_{J0}$ and $\gamma$ from \autoref{tbl:macom-datasheet-values} combined with \eqref{eq:diode-cv-equation}. A typical curve is also provided, as well as the ranges of $C_J$ at discrete voltages recorded on the datasheet. 

\begin{table*}[ht]
    \centering
    \caption{Macom MAVR-011020-1411 Datasheet Values}
    \label{tbl:macom-datasheet-values}
    \colorbox{tbl-background}{\begin{tabular}{lclccc}
                                &   Symbol          & Test Conditions                   & Min.  & Typ.   & Max. \\\midrule
        Junction capacitance    & $C_{J0}$ (pF)     & \SI{1}{\MHz}, \SI{0}{\volt}       & 0.19  & ---    & 0.275 \\
                                &                   & \SI{1}{\MHz}, \SI{4}{\volt}       & 0.065 & ---    & 0.097 \\
                                &                   & \SI{1}{\MHz}, \SI{15}{\volt}      & 0.025 & ---    & 0.048 \\
        \addlinespace[1ex]
        C-V curve slope         & $\gamma$          & \SIrange{2}{12}{\volt}            & ---   & 0.9    & 1.1 \\
    \end{tabular}}
\end{table*}

\begin{figure*}[htbp!]
    \includegraphics[width=6in,center]{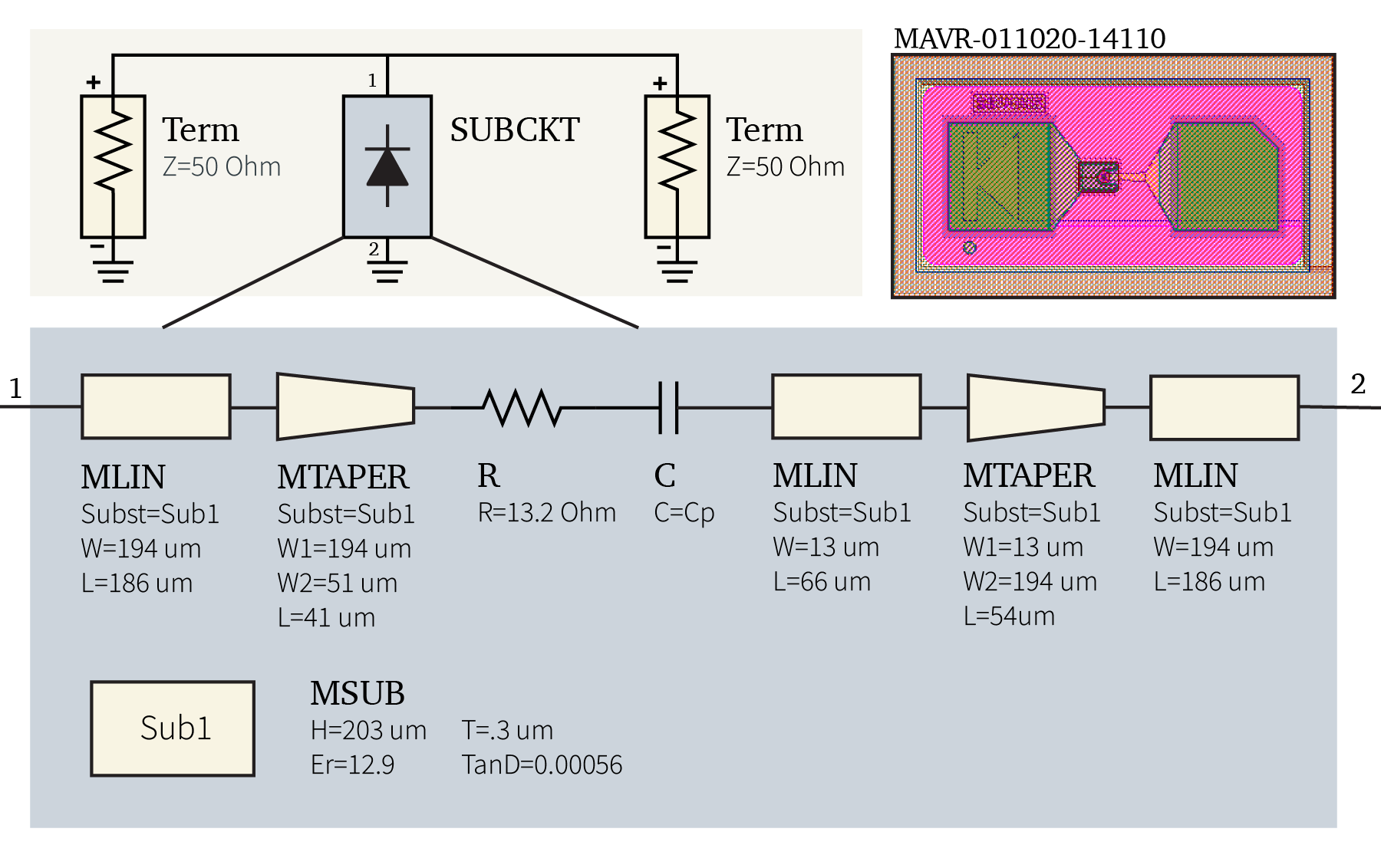}
    \caption{Microstrip transmission line model provided by Macom for the MAVR-001020-1411 varactor diode \cite{mavr011020spice} and implemented in PathWave ADS. A two-port S-parameter network is solved (top left) with a sub-circuit (SUBCKT) representation of the diode (bottom) which includes transmission lines which can be seen in a top-down view of the diode (top right).}
    \label{fig:diode-ads-model}
\end{figure*}

The junction capacitance $C_{J}$ describes a capacitance that occurs at an extremely small junction in the center of a packaged diode; the actual impedance response of the diode must include the effect of transmission lines on the diode and other elements of the package. The flip-chip nature of this particular diode minimizes many of the large parasitic inductances associated with wire-bonds and ceramic or plastic package leads, but for a diode with such a small junction capacitance, even capacitive coupling between pads on the wafer and the inductive transmission lines have a large effect. Macom provides a transmission line model for the diode package \cite{mavr011020spice}, shown in \autoref{fig:diode-ads-model}, from which it is possible to derive an accurate equivalent circuit for the diode. The transmission lines tapers from bond pads to diode are modeled as a series of components in a sub-circuit using PathWave Advanced Design System (ADS). The diode itself, between the transmission lines, is modeled simply by a capacitor which can be parametrically varied based on the junction capacitance and a series resistor, set to the datasheet value of \SI{13.2}{\ohm}. This sub-circuit is modeled in a \SI{50}{\ohm}-terminated S-parameter simulation, from which the equivalent diode impedance may be calculated using (\eqref{eq:Zeq-from-S11}).

\begin{figure*}[htbp!]
    \includegraphics[width=6in,center]{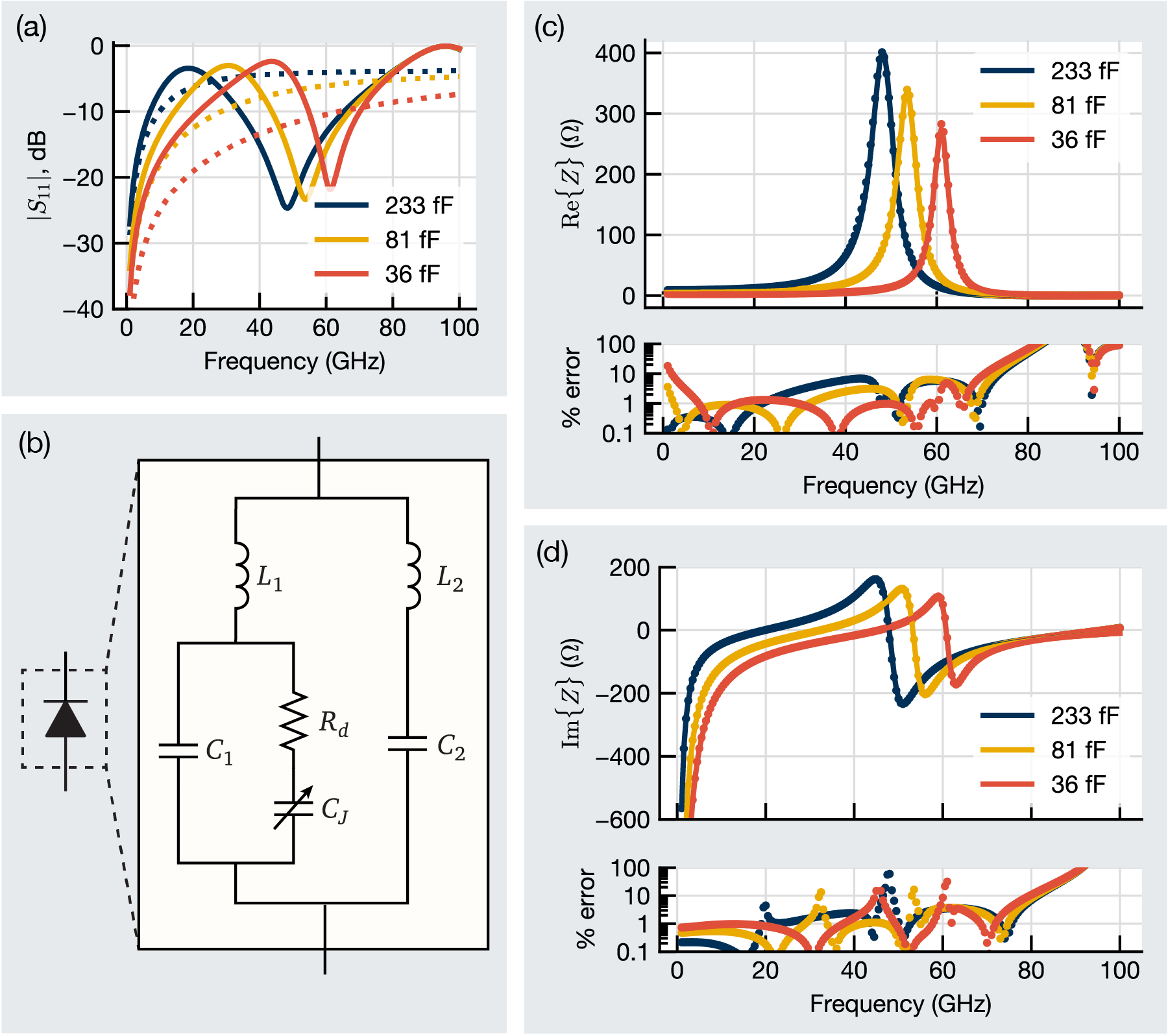}
    \caption{(a) Reflection coefficient obtained from the ADS simulation described in \autoref{fig:diode-ads-model} is shown in solid for three different junction capacitances. The dashed lines show the reflection coefficient for the simple small-signal model of \autoref{fig:diode-small-signal}. (b) A lumped element model is obtained for the flip-chip varactor diode which includes package parasitics. (c) Fit of the lumped element model resistance (lines) of (b) to the resistance obtained by ADS simulations (dots). (d) Same as (c), but for the reactance.}
    \label{fig:diode-ads-extraction}
\end{figure*}

The objective of this exercise is to develop a lumped component model for the diode which can easily be incorporated into the equivalent circuit model of the surface, or more generally, be used as a sub-component in an analytic model or a fast circuit solver such as SPICE. To ensure that the lumped model matches the ADS model over the entire capacitance range of the diode, three separate ADS simulations are performed --- two at the limits of the range (\SI{36}{\fF} and \SI{233}{\fF}) --- and one between the two at \SI{81}{\fF}. The reflection coefficient magnitude $S_{11}$ obtained from these simulations is shown in \autoref{fig:diode-ads-extraction}(a), and is compared to the expected reflection coefficient for the series RC circuit absent the transmission lines to illustrate the importance of this analysis, as the results from the ADS model diverge dramatically above \SI{1}{GHz}.

The values for the equivalent circuit are solved by fitting to the measured impedance of all three models simultaneously. Every component value should be identical in the three cases except for for junction capacitance, which is already known for each case. Inspection of the impedance obtained from the ADS simulations, shown in \autoref{fig:diode-ads-extraction}(c-d) offers some insight into the topology of the circuit. The reactance curve shows two zeros, the first below \SI{45}{\GHz} and which varies with the junction capacitance. The second zero is around \SI{90}{\GHz} and is constant between the cases. Between these two zeros is a pole, the effect of which is suppressed by resistive loss in the diode. These facts suggest the circuit is nominally a Foster's circuit of the second type, with two series LC circuits in parallel. Initial attempts at fitting to this four-element circuit reveal that an additional capacitor in parallel with the junction capacitance and resistance is necessary to move the pole to the correct frequency; the final equivalent circuit is shown in \autoref{fig:diode-ads-extraction}(b). Excellent agreement between ADS and the ECM over the entire frequency range for all three junction capacitances is shown for the resistance in \autoref{fig:diode-ads-extraction}(c) and the reactance in \autoref{fig:diode-ads-extraction}(d). The equivalent circuit values obtained were $L_1 = \SI{0.275}{\nH}$, $C_1 = \SI{9.19}{\fF}$, $L_2 = \SI{0.052}{\nH}$, and $C_2 = \SI{38.3}{\fF}$. 

\begin{figure}[ht!]
    \includegraphics[width=3.5in,center]{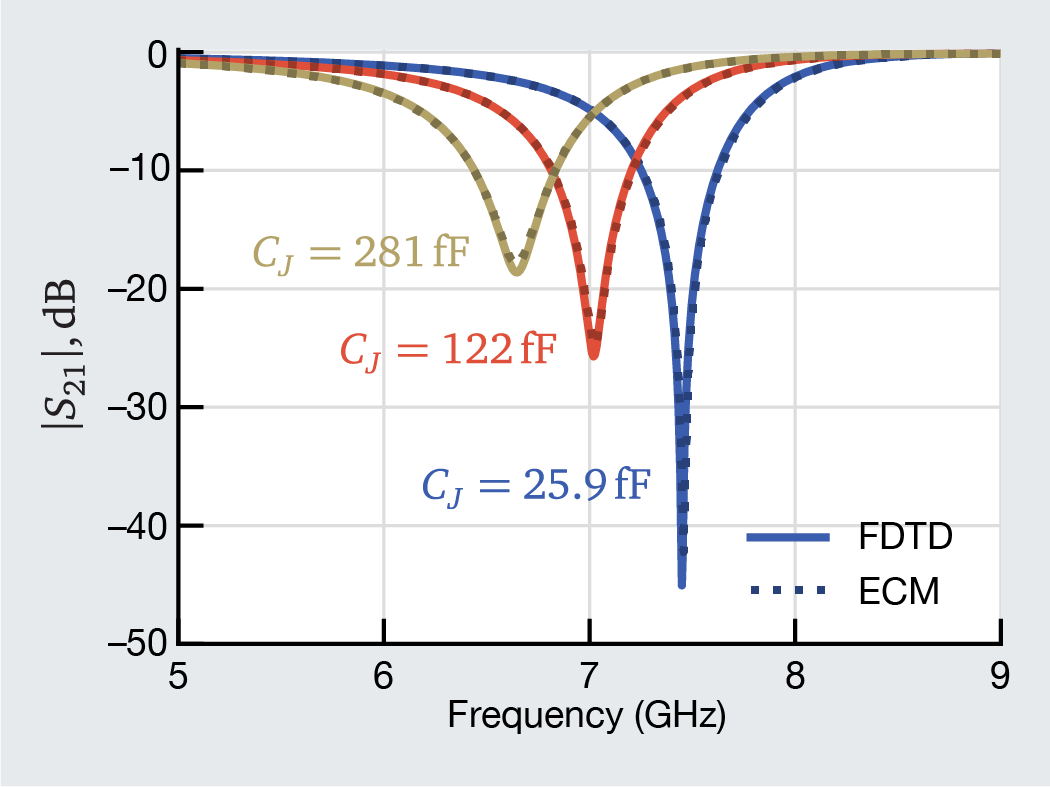}
    \caption{Comparison between FDTD-SPICE results and ECM results for the dogbone surface for three different representative capacitances, incorporating the varactor small-signal representation given in \autoref{fig:diode-ads-extraction}.}
    \label{fig:full-model-ecm-fdtd}
\end{figure}

The final step in model development is confirming that the combination of lumped diode model just obtained and the surface ECM derived earlier suitably predicts the scattering from the surface for any desired junction capacitance, and consequently any applied bias voltage. The lumped circuit model can be divided into two parallel branches
\begin{align}
    Z_1 &= \jw L_1 + \frac{1}{\jw} \left(C_1 + \frac{C_J}{1 + \jw R_d C_J}\right)^{-1} \\
    Z_2 &= \jw L_2 + \frac{1}{\jw C_2}
\end{align}
which are combined with the parallel combination to yield the total diode impedance
\begin{equation}
    Z_L = \left(\frac{1}{Z_1} + \frac{1}{Z_2}\right)^{-1}
\end{equation}
This diode model is embedded into the analytical ECM and used to compute transmission and reflection from the loaded dogbgone surface for a few different junction capacitance values. The scattering parameters are then verified against an FDTD-SPICE simulation with a SPICE netlist containing the same lumped circuit model. A comparison in the resonance curves between FDTD and ECM may be found in \autoref{fig:full-model-ecm-fdtd}, showing good agreement.

\section{Biasing of varactor-loaded surface}

The necessity of biasing circuits poses another challenge for the design of reconfigurable metasurfaces, as biasing traces have a tendency to couple incident electromagnetic energy which, in addition to potentially de-tuning the fundamental resonance, can also create spurious resonances.

Tunable band-pass surfaces by their nature tend to have highly interconnected elements which provides an obvious source for to at least one of the supply voltages required \cite{lin2018}; high-impedance surfaces which are metal-backed have an additional interconnected voltage reference without any further modification \cite{sleasman2023}. Band-stop filters, such as the end-loaded dipole considered here, have disconnected elements and so two bias lines---a ground reference and a bias voltage---must be supplied to each element without perturbing the designed resonance.

\begin{figure*}[htb!]
    \includegraphics[width=6in,center]{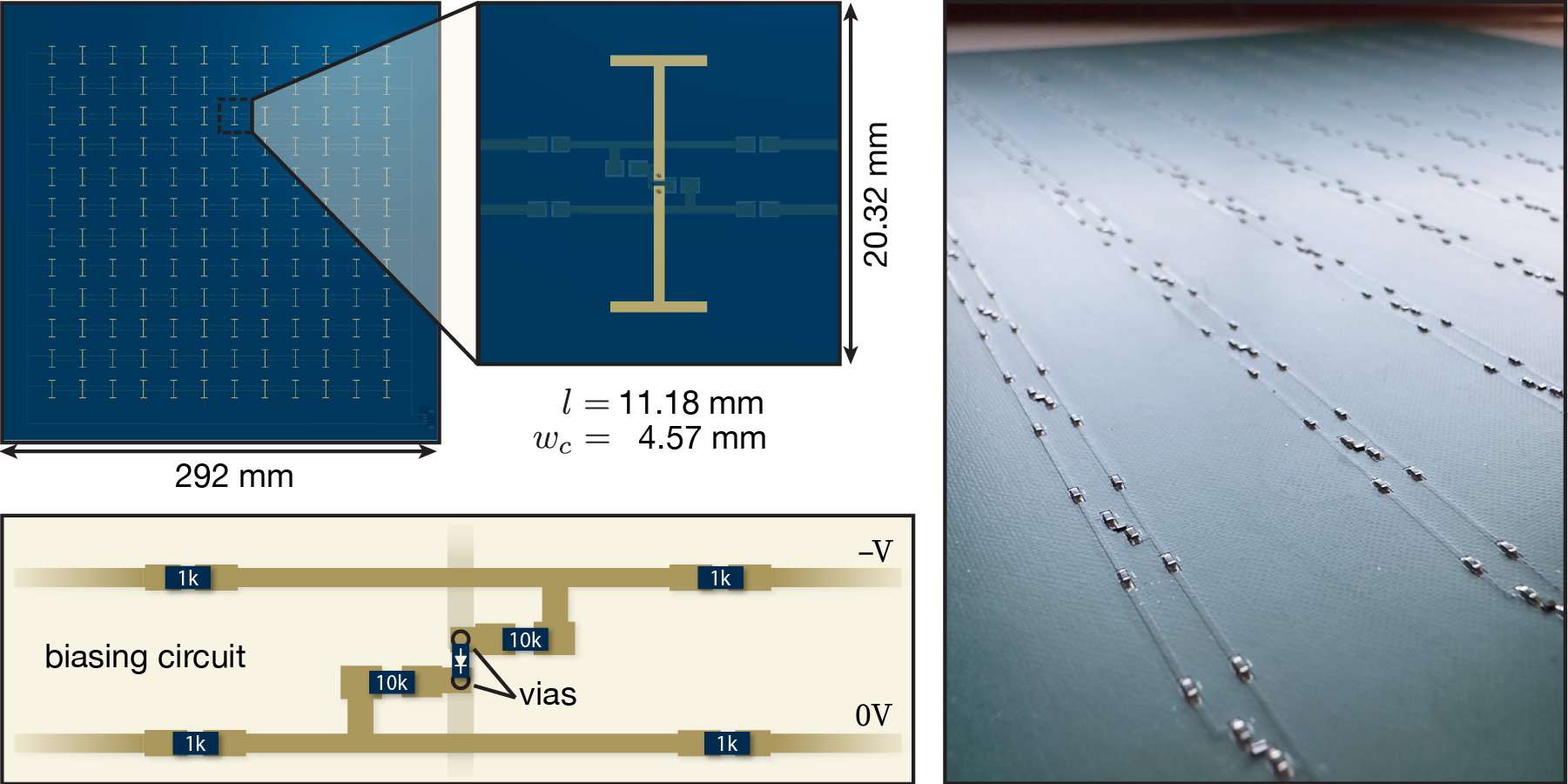}
    \caption{Biasing design for reconfigurable dipole surface.}
    \label{fig:biasing-design}
\end{figure*}

The diode has already been accommodated in the dogbone metasurface by splitting the dipole element at its center and routing current to the substrate backside, where the varactor is placed. The backside of the substrate is therefore a natural location for the bias lines. By running the lines orthogonal to the dipole element, coupling of RF energy into the lines is reduced. Further coupling reduction is achieved by periodically inserting resistors---two \SI{1}{\kohm} resistors per unit cell---along each bias line, which is possible for this application due to the low reverse saturation current of the diodes in use ($<\SI{100}{\nA}$). In addition, one \SI{10}{\kohm} is placed as close as possible to the diode pads to reduce any capacitive coupling from the bias traces. The layout of these elements is illustrated in \autoref{fig:biasing-design}. The bias lines are collected at the edge of the test article and voltage supplied through a DC connector.

Despite the efforts to mitigate the impact of the biasing lines to the electromagnetic response of the surface, some coupling does occur, very slightly extending the electrical length of the dipole. These effects can however be fully captured in the ECM, provided that the bias lines are included in the FDTD model used to create the ECM.

\section{Model validation through measurement}

\begin{figure*}[htb!]
    \includegraphics[width=6in,center]{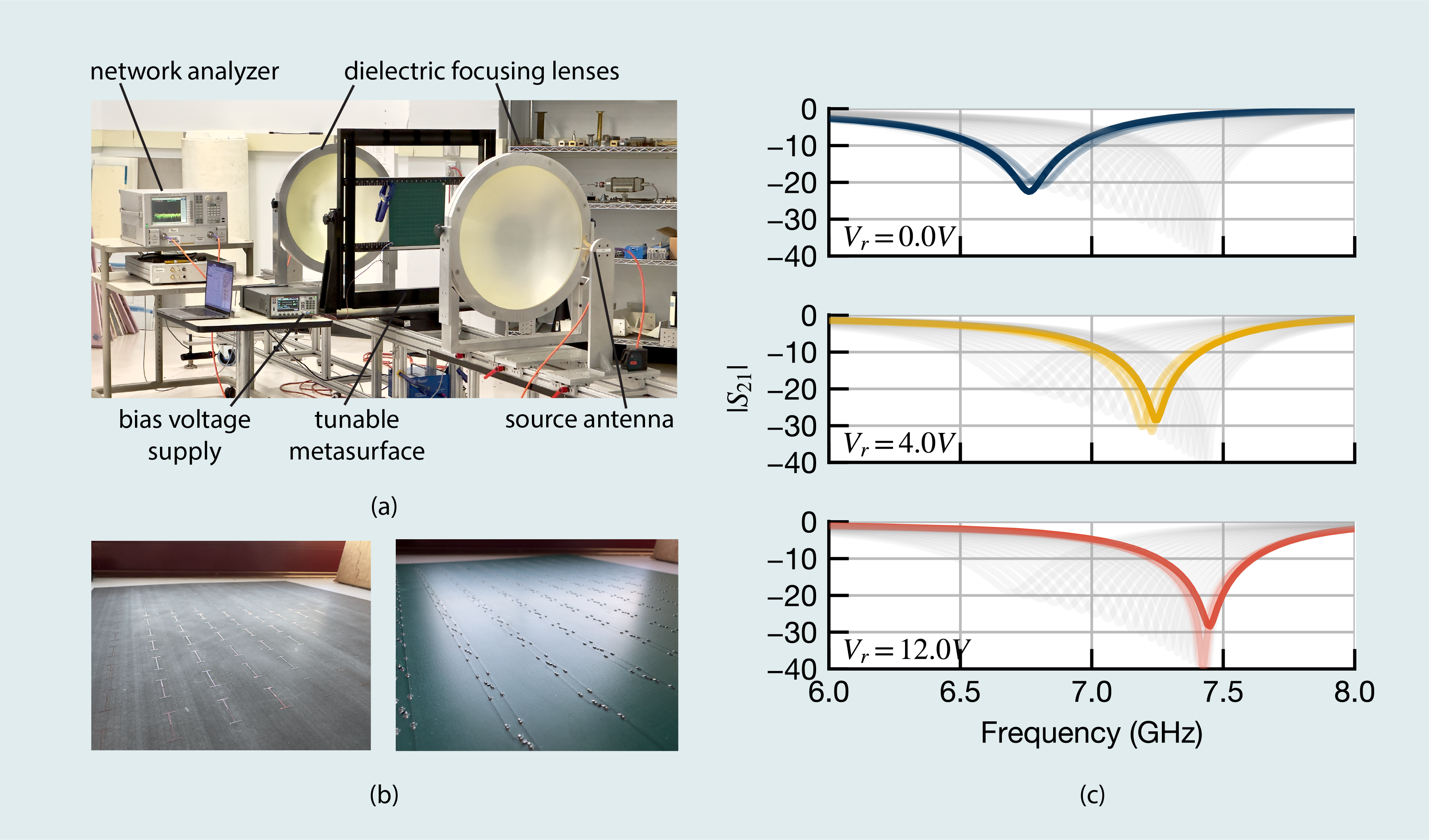}
    \caption{(a) Free-space focused beam measurement setup for obtains transmission data for the tunable metasurface. (b) The front and back (left to right) faces of the fabricated test article. (c) Comparison of measured transmission for bias conditions $V_r$ of 0V, 4V, and 12V against predictions obtained from the FDTD-SPICE simulations. The light gray lines trace out the simulated transmission curves for the entire capacitance range of the diode, while the lightly shaded lines highlight those which correspond to the range of capacitance values given by the datasheet for $V_r$. The dark line is the focused beam measurement at bias voltage $V_r$.}
    \label{fig:msmt-model-agreement}
\end{figure*}

A test article was designed and fabricated with the dimensions and materials listed in \autoref{tbl:design-parameters} and characterized in GTRI's free-space focused beam system as described in \cite{howard2022b}. Agreement between the model and measurement can be shown in \autoref{fig:msmt-model-agreement}, with the measured curves fitting within the simulation results prescribed by the range of junction capacitance values obtained from the datasheet.

\begin{table}
    \centering
    \caption{Design parameters of fabricated dogbone surface}
    \label{tbl:design-parameters}
    \colorbox{tbl-background}{\begin{tabularx}{\linewidth}{lcrl}
                                        & Variable           & Value & Unit\\\midrule
        \tablesectionheader{Surface Dimensions}\\
        Unit cell period                 & $\Lambda$         & 20.32 & mm \\
        Dipole length                    & $l$               & 11.40 & mm \\
        Trace width                      & $w$               & 0.508 & mm \\
        End-loading width                & $w_c$             & 4.572 & mm \\
        Biasing gap spacing              & $g$               & 0.254 & mm \\
        \addlinespace[2ex]
        \tablesectionheader{Substrate (AGC TSM-DS3M)}\\
        Dielectric constant             &  $\epsilon_r'$    & 3.0   & --- \\
        Loss tangent (10 GHz)           &  $\tan\delta$     & 0.0014 & --- \\
        Thickness                       &  $t_s$            & 1.524 & mm \\
    \end{tabularx}}
\end{table}


\section{Conclusion}

From a series of three full-wave solutions, an equivalent circuit model is fully determined, which serves as a robust representation of the relationship between fields in a periodic cell and can thereafter be used to describe the impedance expected from any given circuit in the SPICE network. The equivalent circuit model represents a complete characterization of the electromagnetic portion of the structure, which means that included sub-cell circuit features can be varied without triggering another re-solve in the full-wave solver. In principal, once the equivalent circuit model has been determined with the approach described here, no further full-wave models are necessary, which has the potential for huge computational savings in the design of reconfigurable, time-varying, and non-Foster surfaces, where the domain of possible circuit impedance may be quite large. While not presented here, we expect that nonlinear circuits coupled to EM structures can be simulated through this approach as well through transient or harmonic balance simulation results from SPICE or ADS where the ECM for the metasurface is included in the SPICE network.

It is also apparent that sub-wavelength parasitics in any lumped components must be considered to achieve model-measurement agreement for reconfigurable metasurfaces. The small-signal models provided by many datasheets are typically characterized at low frequencies (such as \SI{1}{\MHz}) and are unreliable in predicting the parasitics at microwave frequencies. A more robust small-signal model can be obtained through electromagnetic modeling, such as through ADS as in this work, or through microwave bench-top characterization. Either of these approaches greatly increase the complexity and labor requirements of designing reconfigurable metasurfaces, but cannot be discarded without great impact to model performance.

We also note the utility of this work as a validation case for hybrid co-simulator implementations. Much of the research documented here was initially performed to test the self-consistency of the FDTD-SPICE algorithm implementation. Historically, the best available method for verifying the performance of a hybrid solver has been to compare simulated results against some measurement; however, the validity of any results obtained from hybrid full-wave/circuit solvers is difficult to assess due to a lack of suitable analytic test cases and compounding uncertainties associated with component tolerances and measurement errors. Under the assumption that the results generated from a full-wave solver -- not a hybrid solver -- with different material properties at a circuit port location should be consistent with the results obtained from a hybrid solver where the circuit contains lumped component representations of the same material properties, such as representing a cell of lossy material in FDTD as a resistor in the SPICE network, then confidence in the hybrid solver's implementation may be improved. The excellent agreement demonstrated among the ECM, FDTD-SPICE, and measured results further bolsters the validity of the ECM and FDTD-SPICE approaches.

\section*{Acknowledgment}

The authors acknowledge and thank Dr. Ryan Westafer and Dr. Jonathan Andreasen for their efforts towards implementing the FDTD-SPICE algorithm used in this work. 

\nocite{*}
\bibliographystyle{jpier}
\bibliography{references}

\end{document}